\pdfoutput=1
\RequirePackage{fix-cm}
\documentclass[british]{article}

\usepackage{spconf} 

\usepackage[utf8]{inputenc}
\usepackage[T1]{fontenc}
\usepackage[british]{babel}

\usepackage{pifont}

\usepackage[pdftex]{graphicx}
\usepackage[table,xcdraw]{xcolor}
\graphicspath{{./figures/}}
\DeclareGraphicsExtensions{.pdf,.png,.jpeg,.jpg}

\usepackage[cmex10]{amsmath} 
\usepackage{amssymb}
\usepackage{bm} 
\usepackage{esint} 
\usepackage{commath} 
\usepackage{units} 

\usepackage{subcaption} 

\usepackage{booktabs} 
\usepackage{array} 

\usepackage{multirow} 
\usepackage{rotating} 

\usepackage[inline]{enumitem} 

\usepackage{comment}

\usepackage{tikz}
\usetikzlibrary{positioning}
\usetikzlibrary{arrows}
\usetikzlibrary{shapes}
\usepackage[unicode=true,pdfusetitle,%
 pdfauthor={Pablo Pérez Zarazaga, Gustav Eje Henter, Zofia Malisz},%
 pdfkeywords={whispered speech, WAD, human-in-the-loop, autonomous sensory meridian response},%
 bookmarks=true,bookmarksnumbered=true,bookmarksopen=false,%
 breaklinks=true,pdfborder={0 0 0},backref=false,colorlinks=true,citecolor=blue]{hyperref}
\usepackage[capitalise]{cleveref} 

\renewcommand{\mid}{\,\ifnum\currentgrouptype=16 \middle\fi|\,}

\ninept 

\let\oldmarginpar\marginpar
\renewcommand\marginpar[1]{\-\oldmarginpar[\raggedleft\footnotesize #1]%
{\raggedright\footnotesize #1}}

\title{A processing framework to access large quantities of whispered speech found in ASMR} 
\makeatletter
\name{Pablo Pérez Zarazaga, Gustav Eje Henter, Zofia Malisz\thanks{This work was supported by the Swedish Research Council grant no.\ 2017-02861 ``Multimodal encoding of prosodic prominence in voiced and whispered speech'' and partially supported by the Wallenberg AI, Autonomous Systems and Software Program (WASP) funded by the Knut and Alice Wallenberg Foundation.} 
}
\address{Division of Speech, Music and Hearing, KTH Royal Institute of Technology, Stockholm, Sweden}

\begin{document}
\maketitle
\begin{abstract}
Whispering is a ubiquitous mode of communication that humans use daily. Despite this, whispered speech has been poorly served by existing speech technology due to a shortage of resources and processing methodology. To remedy this, this paper provides a processing framework that enables access to large and unique data of high-quality whispered speech. We obtain the data from recordings submitted to online platforms as part of the ASMR media-cultural phenomenon. We describe our processing pipeline and a method for improved whispered activity detection (WAD) in the ASMR data.
To efficiently obtain labelled, clean whispered speech, we complement the automatic WAD by using Edyson, a bulk audio-annotation tool with human-in-the-loop. We also tackle a problem particular to ASMR: separation of whisper from other acoustic triggers present in the genre. We show that the proposed WAD and the efficient labelling allows to build extensively augmented data and train a classifier that extracts clean whisper segments from ASMR audio.

Our large and growing dataset enables whisper-capable, data-driven speech technology and linguistic analysis. It also opens opportunities in e.g.\ HCI as a resource that may elicit emotional, psychological and neuro-physiological responses in the listener.

\end{abstract}
\begin{keywords}
Whispered speech, WAD, human-in-the-loop, autonomous sensory meridian response
\end{keywords}
\section{Introduction}
\label{sec:intro}

Whispered speech is a very common mode of vocal communication that comes naturally to humans. It is often used to reduce the audibility of the speech signal~\cite{tartter1989whats} for privacy of 
information. Whisper is also associated with situations of intimacy and speakers may use it to elicit emotion and relaxation. The latter effect contributes to e.g.\ the popularity of the so-called autonomous sensory meridian response (ASMR) genre on streaming platforms.

Despite the communicative ubiquity and evocative power of whispering, we do not see many implementations of whisper modes in technology such as in voice assistants. Only recently have leading companies begun to provide
voice assistants that recognise whispered commands and provide responses in whisper~\cite{cotescu2020voice, rekimoto2022dualvoice}. One of the reasons being that whispered speech datasets are both rare and small, limiting the development of data-driven methods for whisper in speech technologies. 
Studies have also shown a related problem, where existing systems with whispered interaction are judged as ``creepy''~\cite{parviainen2020experiential}. Owing to the lack of whispered speech resources, such systems use phonated-to-whisper speech conversion which possibly contributes to the unconvincing signal generation.
It is, therefore, clear that there is a need for more whisper data sources and for processing methods of whispered material.

In this work, we set out to make more whispered speech data available to speech sciences and technology. We have noticed that very large amounts of whisper data can be found in ASMR videos uploaded to streaming platforms such as YouTube and Twitch. ASMR (the name is pseudo-scientific\footnote{\href{https://www.smithsonianmag.com/science-nature/researchers-begin-gently-probe-science-behind-asmr-180962550/}{https://www.smithsonianmag.com/science-nature/researchers-begin-gently-probe-science-behind-asmr-180962550/}}) is an increasingly popular phenomenon that uses auditory and visual cues to evoke sensory and other effects in listeners. 
Whispered speech is a predominant trigger of ASMR effects~\cite{andersen2015now}, but it is typically mixed with a variety of other auditory triggers. To access the wealth of whisper embedded in found ASMR data, it is necessary to separate the whispered segments from the other signals typical for the genre. Considering the amount of data available, this can become a tedious and time consuming task if performed manually. At the same time, conventional signal processing tools (e.g.\ denoising) that can speed up such processing, work poorly on whispered speech due to the noise-like nature of the signal~\cite{wenndt2002study}.





In this work, we deliver a processing framework that provides access to large amounts of whispered speech data included in ASMR recordings.\footnote{The pre-trained models are available through the following link:\\{}
\href{https://github.com/perezpoz/CleanWhisperDetection}{https://github.com/perezpoz/CleanWhisperDetection}} 
We also separate the whispered signal from other signals typically present in the genre 
by leveraging acoustic features found to be useful in WAD, and improve upon the WAD state-of-the-art by including recurrent neural networks (RNNs) to model time dependencies. To raise the efficiency of processing very large amounts of data, we use Edyson~\cite{fallgren2019annotate}, an application for semi-automatic labelling of audio data. Finally, we use data augmentation to fine-tune a \emph{clean whisper speech detector} (CWAD) for the specific speech style and acoustic triggers present in a set of ASMR recordings. We present this approach as a generalisable method to be used in similar data scenarios.


\section{Background}
\label{sec:background}

Whispered speech is characterised by the lack of vibration of the vocal folds. This causes an absence of fundamental frequency in the speech signal, which, in turn, is visible as an energy reduction in the lower frequencies of the spectrum~\cite{tartter1989whats, jovivcic1998formant}. While other features like the positions of the formants do not change much from whispered to phonated speech,
the lack of $f_0$ complicates the use of speech processing methods adapted to phonated speech. 


Currently available whispered speech resources usually comprise studio recordings of written sentences being whispered (e.g.\ the wTIMIT, wMRT~\cite{lim2011computational} and CHAINS~\cite{cummins2006chains} datasets). However, these datasets contain a relatively small amounts of whispered speech that add up to a total of approx.\ 30 hours. Additionally, custom recorded data of text prompts are likely to lack realistic prosody, and other linguistic variability present in whispered speech outside the lab~\cite{batliner1995can}. 

Rather than lab-recorded corpora, speech processing in general has recently seen substantial advances from leveraging found data resources instead \cite{panayotov2015librispeech,ito2017lj,radford2022robust}. Doing so sometimes necessitates novel processing methods that combine signal processing, machine learning, and human annotation in order to efficiently extract the portions of interest from the found data, e.g.\ \cite{szekely2019casting}.
%

For these reasons, we propose to leverage ASMR recordings as a source of whispered speech. ASMR is a phenomenon that has grown in popularity on streaming platforms, e.g. there are currently 5.2 million ASMR videos on YouTube.
%
The size of the data available for download is approx. hundreds of thousands of hours. 
By recording whispered speech close to a high-sensitivity microphone, ASMR performers aim to transmit physiological responses such as frisson, particularly involving the pilomotor reflex (``goosebumps''), as well as feelings of comfort, intimacy or relaxation~\cite{barratt2015autonomous, del2016autonomous, smith2017examination, poerio2018more}. 
ASMR contains a wide variety of spontaneous whisper, usually mixed with other triggers, for example, lip and tongue smacking and clicking or rubbing the microphone with different materials, free-styled for the purpose of evoking ASMR effects. 



 To excerpt clean speech from large datasets such as ASMR recordings, voice activity detection (VAD) is typically used - a system that recognises which parts of the recording are clean speech.
However, off-the-shelf VAD is generally not applicable to whisper, due to the noise-like quality of the signal.
WAD is a special case of VAD in which the system decides whether the signal contains whispered speech. It is then possible to define two types of WAD: \begin{enumerate*}
\item Distinguishing between phonated and whispered speech, and 
\item detecting if the processed noisy signal contains whispered speech
\end{enumerate*}. 

Whispered speech has reduced audibility in general, while whisper found in ASMR, typically recorded with high sensitivity microphones, poses additional challenges to speech processing methods, particularly whenever other triggers are present. In this work, we propose a third mode of whisper activity detection: separating clean whispered speech from speech affected by noise. We will refer to this method as CWAD. 

Different types of speech input call for different processing approaches. To process and analyse phonated vs.\ whispered speech signal, it is important to distinguish between proper tools.
Several solutions derived from VAD using deep learning have been studied to separate phonated from whispered speech~\cite{raeesy2018lstm, ashihara2019neural,naini2020whisper} but the similarity between whispered speech and non-speech noise, and its lower SNR values produced by the lack of energy in the lower frequencies, lower the usefulness of these methods for detecting whispers within noise. Therefore, WAD systems that detect whispered speech in noise usually rely on custom features tailored for this task. In particular, linear prediction features such as relative spectral filtering on perceptual linear prediction (RASTA-PLP)~\cite{sarria2013whispered,markovic2017recognition} and long-term features like spectral modulation~\cite{ meenakshi2015robust,wang2016whispered} have been found 
useful for distinguishing whispered speech from noise. Our work shows how a combination of feature design and deep learning can improve on existing WAD methods. Additionally, these methods can also be tuned for CWAD to 
label clean whisper data.


\section{Data processing framework}
\label{sec:experiments}


In this paper, we introduce a processing framework that allows separating clean whispered speech from noisy segments in unknown data. The overall approach can be used to automatically label data from unseen speech and noise data in many different scenarios, incl. the use case of gathering whispered speech data from ASMR recordings.
We also provide a pre-trained CWAD system that handles this source of whisper data. The framework is illustrated in Fig.~\ref{fig:FW_flowchart} and detailed below:

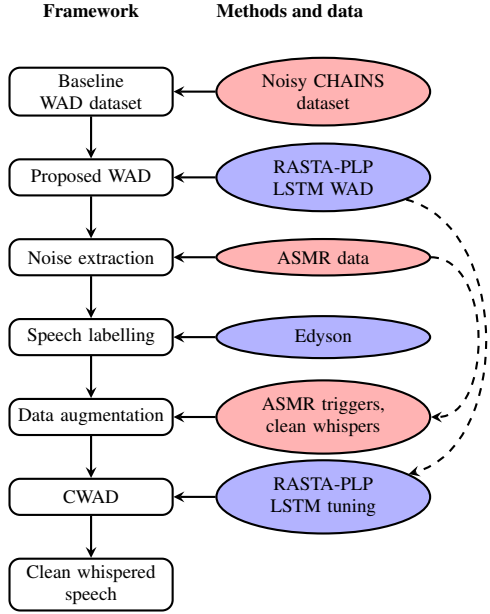
\begin{figure}[t]
    \centering
    \begin{tikzpicture}[auto, node distance = 7mm and 7mm,thick,scale=0.8, every node/.style={transform shape}]
        \node(input) [draw, rectangle, rounded corners, text width=2.5cm, align = center, minimum height = 0.6cm] {Baseline WAD dataset};
        \node(title) [above =of input, minimum width=2.5cm]{\textbf{Framework}};
        \node(contr) [right =of title, minimum width=2.5cm]{\textbf{Methods and data}};
        \node(WAD) [draw, rectangle, rounded corners, below =of input, text width=2.5cm, align = center, minimum height = 0.6cm] {Proposed WAD};
        \node(NExt) [draw, rectangle, rounded corners, below =of WAD, text width=2.5cm, align = center, minimum height = 0.6cm] {Noise extraction};
        \node(Edy) [draw, rectangle, rounded corners, below =of NExt, text width=2.5cm, align = center, minimum height = 0.6cm] {Speech labelling};
        \node(DAug) [draw, rectangle, rounded corners, below =of Edy, text width=2.5cm, align = center, minimum height = 0.6cm] {Data augmentation};
        \node(Clean) [draw, rectangle, rounded corners, below =of DAug, text width=2.5cm, align = center, minimum height = 0.6cm] {CWAD};
        \node(CleanWhsp) [draw, rectangle, rounded corners, below =of Clean, text width=2.5cm, align = center, minimum height = 0.6cm] {Clean whispered speech};
        
        \node(ConIN) [draw, ellipse, fill=red!30!white, right =of input, text width=2.3cm, align = center, minimum height = 0.6cm] {Noisy CHAINS dataset};
        \node(ConWAD) [draw, ellipse, fill=blue!30!white, right =of WAD, text width=2.3cm, align = center, minimum height = 0.6cm] {RASTA-PLP LSTM WAD};
        \node(ConNExt) [draw, ellipse, fill=red!30!white, right =of NExt, text width=2.3cm, align = center, minimum height = 0.6cm] {ASMR data};
        \node(ConEdy) [draw, ellipse, fill=blue!30!white, right =of Edy, text width=2.3cm, align = center, minimum height = 0.6cm] {Edyson};
        \node(ConAug) [draw, ellipse, fill=red!30!white, right =of DAug, text width=2.3cm, align = center, minimum height = 0.6cm] {ASMR triggers, clean whispers};
        \node(ConClean) [draw, ellipse, fill=blue!30!white, right =of Clean, text width=2.3cm, align = center, minimum height = 0.6cm] {RASTA-PLP LSTM tuning};
        
        \draw[-stealth] (input.south) -- (WAD.north) {};
        \draw[-stealth] (WAD.south) -- (NExt.north) {};
        \draw[-stealth] (NExt.south) -- (Edy.north) {};
        \draw[-stealth] (Edy.south) -- (DAug.north) {};
        \draw[-stealth] (DAug.south) -- (Clean.north) {};
        \draw[-stealth] (Clean.south) -- (CleanWhsp.north) {};
        
        \draw[-stealth] (ConIN.west) -- (input.east) {};
        \draw[-stealth] (ConWAD.west) -- (WAD.east) {};
        \draw[-stealth] (ConNExt.west) -- (NExt.east) {};
        \draw[-stealth] (ConEdy.west) -- (Edy.east) {};
        \draw[-stealth] (ConAug.west) -- (DAug.east) {};
        \draw[-stealth] (ConClean.west) -- (Clean.east) {};
        
        \draw[dashed, -stealth] (ConNExt) edge[bend left=90] (ConAug) {};
        \draw[dashed, -stealth] (ConWAD) edge[bend left=75] (ConClean) {};
    \end{tikzpicture}
    \caption{Proposed framework for CWAD. 
    Blue boxes represent implemented methods and red ones represent a specific type of data.}
    \label{fig:FW_flowchart}
    \vspace{-\baselineskip}
\end{figure}


\textbf{1. Proposed WAD:} We build an RNN classifier to detect the beginning and end of any whispered speech segments, including those embedded in noise, consisting of other acoustic triggers or any other noise. We evaluate the performance of the proposed classifier compared to baseline WAD solutions: a standard support vector machine (SVM) RASTA-PLP~\cite{sarria2013whispered,markovic2017recognition} and a multi-layer perceptron (MLP) with RASTA-PLP features. We also compare to a state-of-the-art VAD method~\cite{Bredin2020pyannote,Bredin2021end}. 

\textbf{2. Noise extraction:} We use the best-performing WAD method to extract segments that are 100\% noise (i.e.\ no whispered speech) in the ASMR recordings and save them for data augmentation.

\textbf{3. Labelling with Edyson:} We use Edyson \cite{fallgren2019annotate}, a machine learning-supported human-in-the-loop method, to semi-automatically label clean whisper segments and those corrupted by other acoustic triggers or other noise in ASMR data. The noise-removal in the second step facilitates this process by having had removed non-speech noise from the labelling task.

\textbf{4. Data augmentation:} Subsequently, we generate augmented data combining clean whisper data from the CHAINS corpus and ASMR, and we mix these with ASMR acoustic triggers saved in the second step of the framework.

\textbf{5. Proposed CWAD:} Using the augmented data, the best-performing WAD classifier is fine-tuned to detect which segments of the recorded speech are unaffected by noise and which are corrupted by other acoustic triggers or other noise. This way we obtain clean whispered speech.



\section{Framework details}
\label{sec:method}

In this section, we discuss the different whisper activity detection methods and audio labelling tools we used in the steps of the proposed framework presented in Section~\ref{sec:experiments}.

\subsection{Acoustic features for WAD}

\textbf{RASTA-PLP:} It has been observed that slowly changing features in the speech signal, such as spectral envelope obtained via linear prediction, define whisper and phonated speech well. For the detection of whispered speech in noisy environments, RASTA-PLP features have been used successfully~\cite{sarria2013whispered,markovic2017recognition}. The features are obtained by applying linear prediction to a signal processed by a set of Bark-scaled triangular filters. Usually delta features of first and second order are used to represent the variation of the features over time.

\textbf{Modulation features:} There have also been studies of acoustic feature dynamics in whispered speech. Such signal modulation over time can be captured as variation in long-term energy measurements (LTLEV)~\cite{meenakshi2015robust}, or by estimating the group delay for different frequency bands in the signal spectrum~\cite{wang2016whispered}. These methods, however, require relatively long analysis windows, reducing the time resolution of the WAD. We propose to remove this problem by leveraging the ability to model time dependencies with an RNN and a shorter sequence length instead (cf.\ \ref{subsec:propWAD}).

\subsection{Deep neural networks for WAD}

Conventional WAD uses classifiers such as SVMs or Gaussian Mixture Models (GMMs)~\cite{sarria2013whispered,markovic2017recognition,wang2016whispered} to assign labels (e.g.\ whisper or noise) to the feature vectors of each frame. Deep learning methods, however, are a popular choice in VAD~\cite{Bredin2020pyannote} for distinguishing between phonated and whispered speech~\cite{raeesy2018lstm}.

In this work, by comparing to these standards, we evaluate the utility of neural networks as classifiers for this task. Namely, first, an MLP is used to classify each frame of the RASTA-PLP features. Next, we use a long short-term memory (LSTM)-based~\cite{hochreiter1997long} RNN to integrate information across time. The RNN analyses a sequence of consecutive RASTA-PLP frames, and is therefore able to make decisions that use more information consistently across time.

\subsection{Proposed WAD}
\label{subsec:propWAD}

Using RASTA-PLP and delta features of first and second order as input, three classifiers are trained to detect whisper speech in noisy environments. The selected classifiers are an SVM, a multi-layer perceptron, and an LSTM-based RNN that can take advantage of time dependencies in the signal. Additionally, a pre-trained MLP-based VAD model~\cite{Bredin2020pyannote,Bredin2021end} is included as a state-of-the-art method designed for phonated speech, to indicate how such methods may perform in whispered speech situations.

The input RASTA-PLP features are calculated using an analysis window length of 40~ms and a hop size of 20~ms. We then calculate 19 linear prediction coefficients for each RASTA-filtered frame and extract delta features of first and second order. The resulting 57-dimensional input is used for all the proposed classifiers. The proposed MLP then processes each frame of the input features. This model comprises 3 fully-connected hidden layers of sizes 64, 64 and 8 respectively, all with ReLU activation functions. The output of the network has a sigmoid activation function and results in a one-dimensional value in the range of $(0,1)$, which represents the probability of speech in the current frame.

In turn, the proposed RNN processes a sequence of input features of length 30, which corresponds to an audio segment 0.6~s long. We built the proposed RNN using two unidirectional LSTM layers with 64 hidden units to maintain a low computational complexity in the system. One final fully-connected layer with sigmoid activation processes the output of the last LSTM cell and returns a one-dimensional value in the range $[0,1]$, which represents a probability value for speech being present in the signal. A decision threshold is chosen as the optimum value in the ROC curve~\cite{fawcett2006introduction}.

\textbf{Baseline WAD dataset:} In order to train and evaluate the first step of the framework with the proposed WAD methods, we generate a noisy whispered speech dataset. The clean whispered speech is taken from whispered utterances in the CHAINS dataset~\cite{cummins2006chains}. This dataset contains whisper data from 36 different speakers (16 female and 20 male) with 37 utterances per speaker and a total of 6 hrs of speech. The speakers have an Irish accent, except for four of each gender with a UK accent. Environmental noise is then extracted from the QUT noise dataset~\cite{dean2010qut}, which contains environmental noises such as recordings from a kitchen, a café, the inside of a car with windows open and closed, and a noise source in a reverberant room. Speech from the clean whisper dataset is separated into a training set with 15 male and 12 female speakers, and a test set with 5 male and 4 female speakers. Additionally, we keep 20\% of the training utterances as validation data. Noisy mixtures are generated with 10, 5 and 0~dB SNR, with 50 speech utterances chosen at random for each noise-SNR combination. Each speech utterance is then followed by a silence with the same length as the speech signal, thus resulting in equal parts speech and non-speech data. The speech and noise recordings are resampled to 16~kHz before mixing.


\subsection{Whispered data labelling}

\textbf{ASMR data:} We extract ASMR recordings from online platforms using youtube-dl\footnote{\href{https://youtube-dl.org/}{https://youtube-dl.org/}}. We gather data from 10 speakers, 4 male and 6 female. In order to allow for a robust evaluation of the proposed framework, we focus on recordings of high quality whispered speech, i.e. containing direct whispering into the microphone without continuous background noise and where the metadata suggest that the focus of the recording is the whisper trigger e.g. finger tapping. The data duration per speaker is approximately 6 hours, i.e.: 60 hours in total. The extracted samples are presented in the repository as a set of links to the corresponding video and a download script.

\begin{figure}
    \centering
    \includegraphics[width=.9\linewidth]{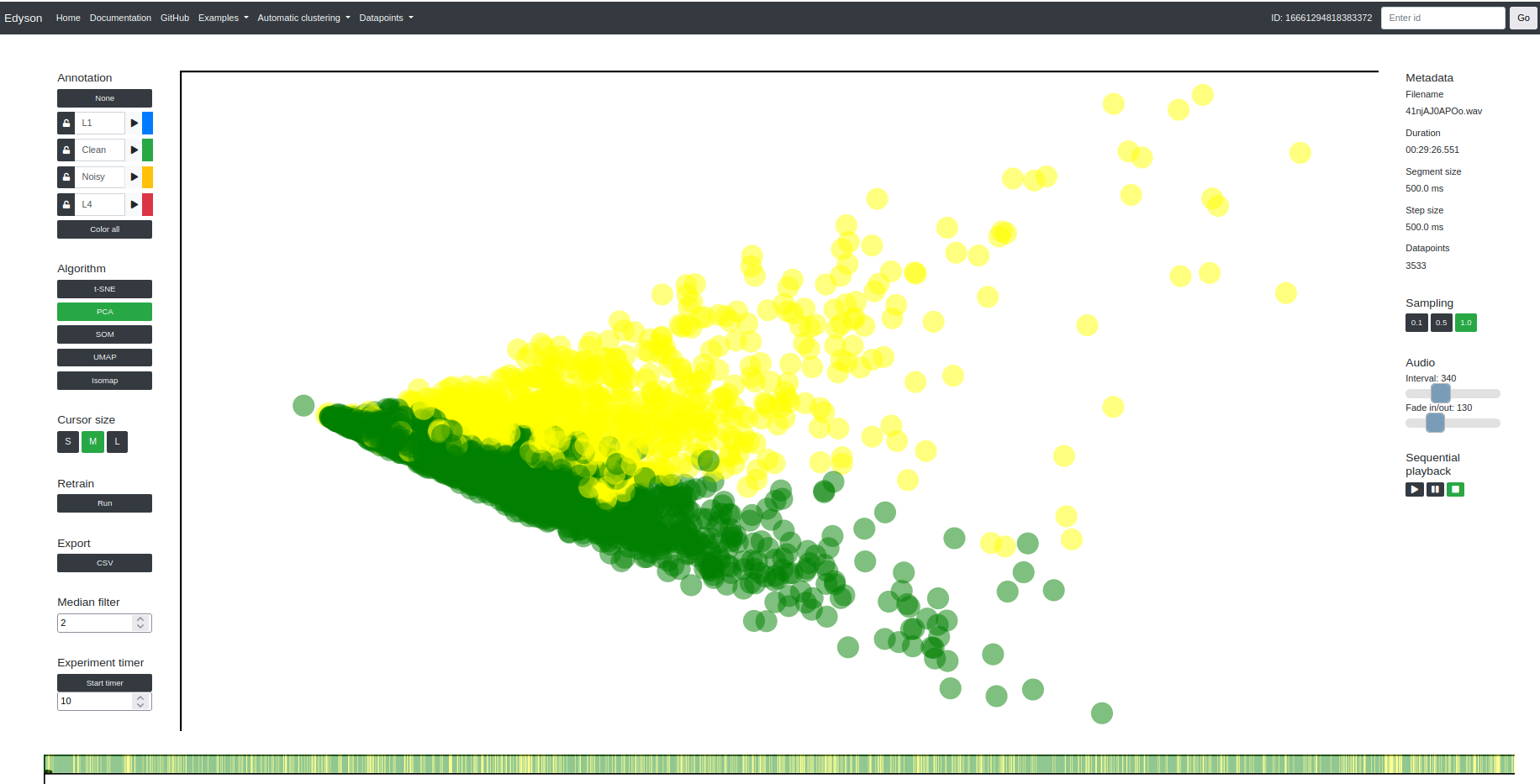}
    \caption{The classification of many whispered speech samples (from ASMR) cast to 2D space using principal component analysis in Edyson. Green data points represent samples labelled as clean speech and yellow ones contain noisy speech.}
    \label{fig:edyson}
    \vspace{-\baselineskip}
\end{figure}

To train data-driven classifiers that adapt to the characteristics of ASMR recordings, it is necessary to provide labels for at least some of the downloaded data. The sheer size of such data, several hundred thousand hours, although in itself an advantage, prohibits manual labelling due to effort and cost. We have previously developed a tool for efficient annotation of large amounts of hitherto unexplored data semi-automatically, allowing to label hours of audio in several minutes. Edyson~\cite{fallgren2019annotate} is a machine learning supported, human-in-the-loop interface that lets the annotator simultaneously assign labels to many instances of similar data. 
Edyson divides the audio signal into snippets of fixed length between 100 and 1000~ms, and extracts spectral features of speech, such as MFCCs~\cite{Davis1980comparison}. It then applies several dimensionality reduction techniques such as PCA and $t$-SNE and plots the corresponding features in 2D space, as shown in Fig.~\ref{fig:edyson}. The data is presented to the user as points in 2D. The user can listen to multiple segments in the same area and assign a label to all of them with a single click.
Additionally, it is possible to alternate between representations cast by the dimensionality reduction methods, adapting to different characteristics of the data. The labels are then stored with their corresponding timestamps.

Labelling with Edyson has been found to be highly accurate~\cite{fallgren2019annotate}. The time needed for labelling does not scale linearly with the amount of audio being annotated. In practice, using the right input features plays a significant role defining clear boundaries between the different data classes for rapid and easy labelling~\cite{zarazaga2022feature}. We have tested several features to discern whispered from phonated speech and noise, such as multitaper spectra~\cite{babadi2014review} or RASTA-PLP.

\subsection{Augmented data for CWAD}

While we can extract large amounts of whispered speech data from ASMR recordings, it is hard to control the characteristics of these speech signals and the resulting data size. In order to train a CWAD that identifies clean whispered speech, and also to perform a robust evaluation of that method, we create augmented data. First, we combine the clean whispered speech from the ASMR recordings with clean utterances from CHAINS. We then take segments identified as noise by WAD in the ASMR recordings and add them to the clean whisper data. This way, we obtain a set of noisy, labelled audio samples at different SNR levels useful for both training and testing.

\section{Results and discussion}
\label{sec:results}

\begin{table}[t]
    \centering
    \caption{Accuracy and F1-score results for the different whisper activity detection classifiers evaluated on the validation set and test set with different SNR values: (SVM) RASTA-PLP features with SVM~\cite{sarria2013whispered}, (VAD) pre-trained state-of-the-art VAD~\cite{Bredin2020pyannote}, (MLP) custom MLP, (RNN) custom RNN.}
    \begin{tabular}{|c|c|c|c|c|c|}
    \hline
        Metric & Data &  SVM & VAD & MLP & RNN \\
        \hline \hline
        F1 (\%) & validation & 87.63 & - & 89.09 & 94.07\\
        \hline
        Acc (\%) & validation & 89.31 & - & 90.87 & 94.31\\
        \hline
        F1 (\%) & test 10dB & 90.01 & 81.19 & 91.89 & 95.71\\
        \hline
        Acc (\%) & test 10dB & 90.97 & 81.81 & 92.03 & \textbf{95.71} \\
        \hline
        F1 (\%) & test 5dB & 88.38 & 77.13 & 88.06 & 93.69 \\
        \hline
        Acc (\%) & test 5dB & 89.88 & 79.02 & 88.66 & \textbf{93.60}\\
        \hline
        F1 (\%) & test 0dB & 84.33 & 74.89 & 83.30 & 91.34 \\
        \hline
        Acc (\%) & test 0dB & 86.81 & 77.49 & 84.82 & \textbf{91.38}\\
        \hline
    \end{tabular}
    \vspace{-\baselineskip}
    \label{tab:acc_noisy_WAD}
\end{table}

We analyse the performance of the proposed WAD using accuracy and F1-score over the validation and test sets for each SNR level. The results for the WAD step are presented in Table~\ref{tab:acc_noisy_WAD}. We can observe that the pre-trained VAD presents a worse performance than methods that use features adapted to whispered speech. This shows that methods designed for phonated speech do not adapt properly to whisper and different features are required. The best performing method is the RNN classifier, which outperforms the SVM and MLP by about 5\% absolute. 
This method will then be used in the following steps of the framework.

We then apply the RNN classifier on the set of ASMR recordings, thus separating the segments of pure noise triggers and those containing speech. We use Edyson to semi-manually label the corresponding speech signals we obtained as clean or noisy speech. To evaluate the improvement Edyson provides over manual labelling, we timed the Edyson sessions. A listener with several weeks of practice with the tool managed to assign labels to a 30-minute file in 7 minutes, which is at least four times faster than real time. 

\begin{table}[t]
    \centering
    \caption{CWAD confusion matrix containing the fraction of true negatives (TN), false negatives (FN), false positives (FP) and true positives (TP).}
    \vspace{-\baselineskip}
    \begin{tabular}{l|l|c|c|c|c|}
        \multicolumn{2}{c}{}&\multicolumn{2}{c}{SVM}&\multicolumn{2}{c}{RNN}\\
        \cline{3-6}
        \multicolumn{2}{c|}{}&Noisy&Clean&Noisy&Clean\\
        \cline{2-6}
        \multirow{2}{*}{True}& Noisy & TN: $0.34$ & FP: $0.074$& TN: $0.43$ & FP: $0.064$\\
        \cline{2-6}
        & Clean & FN: $0.2$ & TP: $0.38$ & FN: $0.11$ & TP: $0.39$ \\
        \cline{2-6}
    \end{tabular}
    \label{tab:conf_mat}
    \vspace{-\baselineskip}
\end{table}

The results for the CWAD method on ASMR triggers using the proposed RNN classifier are presented in Table~\ref{tab:conf_mat}. The overall accuracy is $82.10\%$. For CWAD, we are not only interested in the detection of clean whisper, but it is arguably more important to reject the segments containing speech corrupted by ASMR triggers and other noise. The trained CWAD system successfully errs on the side of caution, in the sense that its false positive rate results in the smallest value (6.4\%) in Table~\ref{tab:acc_noisy_WAD}.

To the authors' knowledge, there are no other works on CWAD that could be used as a reference for the obtained results. In comparison with the WAD methods in Table~\ref{tab:acc_noisy_WAD}, the quality of the proposed CWAD will need improvement, however, given the challenges this data presents with regards to signal processing e.g.: low overall sound volume and large data size, our framework already significantly improves the efficiency of extracting relevant data.


\section{Conclusions and future work}
\label{sec:conclusion}

Our results show that the proposed WAD outperforms previously studied feature-based methods. Additionally, the presented CWAD well separates clean whispered speech from that affected by other acoustic triggers in ASMR recordings. With just one pass through our proposed processing framework, we have accessed 20 hours of whispered speech from 60 hours of downloaded ASMR data and we are continuously adding more\footnote{Accessible here: \href{https://github.com/perezpoz/CleanWhisperDetection}{https://github.com/perezpoz/CleanWhisperDetection}}. This already corresponds to half the volume of other currently available whispered speech resources, yet obtained in a much more efficient and scalable way. The data we provide access to was found in-the-wild but features a uniquely high recording quality with a variety of whispered speech styles that can now be explored systematically. 

Our future work involves using metadata analysis and machine-assisted human classification 
to categorise the whisper styles present in the genre and provide labels for further supervised learning. We are also working on describing the acoustic characteristics of the most "successful" ASMR audio examples, that is the ones that most efficiently evoke neurophysiological (``tingles'', goosebumps), psychological (relaxation) and emotional (intimacy) effects in the audience. This way we will further make way for studies of communicative whisper of interest to e.g.\ phoneticians and linguists as well as for human-computer interaction research leading to improvements in the acceptability of voice assistants that whisper.

\bibliographystyle{IEEEbib}
\bibliography{refs}

\end{document}